# On the Oxygen Abundance of Neutral Gas in I Zw 18

**Max Pettini**[1] **and Keith Lipman**[2]

[1] Royal Greenwich Observatory, Madingley Road, Cambridge, CB3 0EZ, UK
[2] Institute of Astronomy, Madingley Road, Cambridge, CB3 0HA, UK

**Abstract.** Whether the H II regions of dwarf emission-line galaxies are self-polluted by the nucleosynthetic products of massive stars remains an open question, despite recent claims to the contrary. We show that the finding by Kunth and collaborators that the neutral interstellar medium of I Zw 18 has a metallicity one order of magnitude lower than that of the ionized gas is highly uncertain. The range of possible values of (O/H) admitted by the saturated O I absorption line used in their analysis is very large, spanning a factor of $\sim 1000$. We suggest that future observations of the intrinsically weaker S II $\lambda 1256$ multiplet offer the best prospects for measuring the true degree of metal enrichment of this dwarf galaxy.

## 1. Introduction

In a recent paper Kunth et al. (1994; hereafter KLSV) reported the first measurement of element abundances in the H I gas of I Zw 18, one of the most metal-poor H II galaxies known. From their analysis of the interstellar line of O I $\lambda 1302.1685$ (multiplet 2) recorded with the Goddard High Resolution Spectrograph (GHRS) on board the Hubble Space Telescope (HST) these authors deduced that neutral gas in I Zw 18 has an oxygen abundance of only $\sim 1/1000$ of solar, $\sim 20$ times lower than that of the H II region (Skillman & Kennicutt 1993). Both absorption and emission measurements refer to the line of sight to the brighter NW H II region. This surprising result appears to provide supporting evidence for the earlier suggestion by Kunth & Sargent (1986) that the H II regions of dwarf emission line galaxies are self-polluted in heavy elements, produced by the same massive stars responsible for ionizing the gas, and to fulfil their prediction that: *"Absorption-line studies using background QSOs may reveal that the outer parts of objects like I Zw 18 are composed of pristine gas"*.

The evolutionary status of dwarf galaxies is an important question, of relevance not only to our understanding of galaxy evolution in general but also to the determination of primordial abundances. Consequently, the observation by KLSV has received much attention and has motivated efforts to reconsider the physical processes and timescales for the mixing of interstellar gas (Roy & Kunth 1995; Kunth, Matteucci, & Roy 1995) and to model element ratios in low mass galaxies (Kunth, Matteucci, & Marconi 1995).

In this *Letter* we point out that the HST spectrum obtained by KLSV is actually unsuitable for abundance measurements because the O I line is saturated, and show that the column density of O I cannot be constrained within useful limits. It is *not* our intention to dispute the possibility that H II regions may be self-enriched; we simply argue that this is still an open question which cannot be resolved with existing data. We also suggest future observations which are more likely to yield a reliable determination of the degree of metal enrichment in I Zw 18.

## 2. The Oxygen Abundance in I Zw 18

It is a well known fact (e.g. Gomez-Gonzalez & Lequeux 1975)—and indeed one recognised by KLSV—that once an interstellar line is saturated it no longer serves as a useful measure of the column density of absorbing atoms $N$. The equivalent width and profile of a saturated absorption line depend much more sensitively on the velocity dispersion parameter $b$ ($b = \sqrt{2}\sigma$, where $\sigma$ is the one-dimensional velocity dispersion along the line of sight) than on the column density. This limitation can at times be alleviated when one has access to several resonant transitions of the same ion with sufficiently different values of oscillator strength for a curve-of-growth to be constructed and the value of $b$ deduced. This is however not the case for multiplet 2 of O I; $\lambda 1302.1685$ is the only resonant transition in the multiplet and, with an $f$-value of 0.04887 (Morton 1991), is nearly always saturated, even in gas of low metal abundance (e.g. Pettini, Lipman, & Hunstead 1995). Consequently even *Copernicus* observations of this line in the spectra of bright stars could not be used to measure $N$(O I) reliably, despite the high signal-to-noise ratio and spectral resolution of those data (Zeippen, Seaton, & Morton 1977). Essentially all interstellar determinations to date—including those made with HST—have been based on detections of the intrinsically much weaker ($f = 1.248 \times 10^{-6}$) intersystem line λ1355.5977 (e.g. Cardelli et al. 1993).

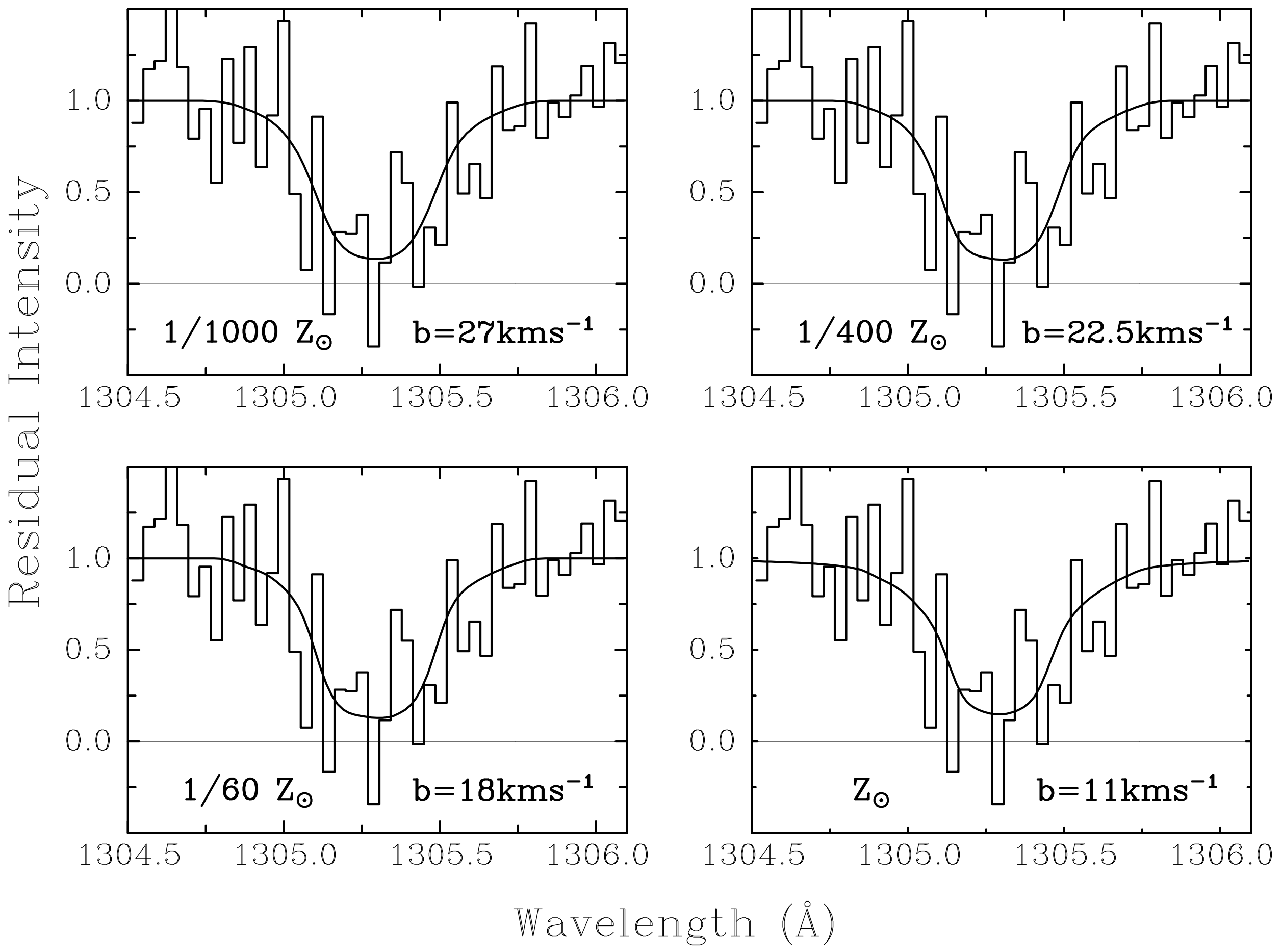

**Fig. 1.** Comparison between the observed profile of the O I λ1302.1685 line in I Zw 18 recorded with the GHRS by Kunth et al. (1994) and theoretical absorption profiles for four possible values of the velocity dispersion parameter $b$. The corresponding values of the oxygen abundance are given in each panel. The saturated λ1302 line cannot be used to constrain the column density of O I within useful limits. The computed profiles have been convolved with the measured Line Spread Function of the GHRS Large Science Aperture which, prior to the HST repair mission, consisted of a narrow core with FWHM = 20 km $s^{-1}$ and broad wings extending to ±70 km $s^{-1}$ (Savage et al. 1993).

Faced with such difficulties, KLSV decided to adopt the value $b = 27$ km $s^{-1}$ from the (unpublished) 21 cm emission observations of I Zw 18 by Viallefond & Lequeux, obtained with the VLA. This is a risky assumption, however. First, the H I column density deduced from Lyman $\alpha$ absorption, $N(\text{H I}) = (3.5 \pm 0.5) \times 10^{21}$ $cm^{-2}$, is only half that seen in 21 cm emission at the same position as the HST aperture, $N(\text{H I}) = 7 \times 10^{21}$ $cm^{-2}$ (KLSV). Presumably half of the gas lies *behind* the stars providing the UV continuum; it is thus quite possible that the 21 cm emission profile covers velocities where no O I absorption occurs and that the $b$-value of O I λ1302 is less than 27 km $s^{-1}$. Second, this large value of $b$ may well result from the presence of multiple velocity components; the precise distribution of O I among the different clouds is the overriding factor determining the column density implied by the line equivalent width. (A value of $b$ as large as 27 km $s^{-1}$ must arise predominantly from bulk, rather than thermal, motions; of course, the thermal component of the velocity dispersion will be 4 times smaller for O than for H).

The point we want to stress is that the profile of the O I absorption line itself provides *no* additional information constraining the choice of $b$ and the resultant estimate of oxygen abundance. This can be seen from Figure 1, where we compare the observed profile of O I λ1302 (re-trieved from the HST data archive) with those computed for a range of values of $b$ and corresponding [O/H]. The theoretical profiles were produced with the *Xvoigt* software package (Mar & Bailey 1995) and convolved with the instrumental resolution of the HST observations. The equivalent width is the same in all four cases illustrated, and was fixed to be $W_\lambda$ = 390 mÅ, the value measured by KLSV. The top left panel shows the fit to the observed profile obtained with KLSV's most probable values: $b = 27$ km s$^{-1}$ and [O/H] $= -3$ (in the usual notation [O/H] $= \log\,(\mathrm{O/H})_{\mathrm{I\ Zw\ 18}} - \log\,(\mathrm{O/H})_\odot$). If $b$ is reduced to 22.5 km s$^{-1}$, we obtain what KLSV regard as their upper limit, [O/H] $= -2.6$ (top right panel of Figure 1). However, if $b = 18$ km s$^{-1}$, the oxygen abundance is the same as that of the NW H II region, [O/H] $= -1.76$ (Skillman & Kennicutt 1993), and an abundance as high as solar is obtained with $b = 11$ km s$^{-1}$ (bottom panels). As can be seen from the Figure, *the absorption profiles corresponding to these four very different solutions are essentially identical.* Similar considerations apply to Si II λ1304.3702 also covered by the GHRS spectrum obtained by KLSV.

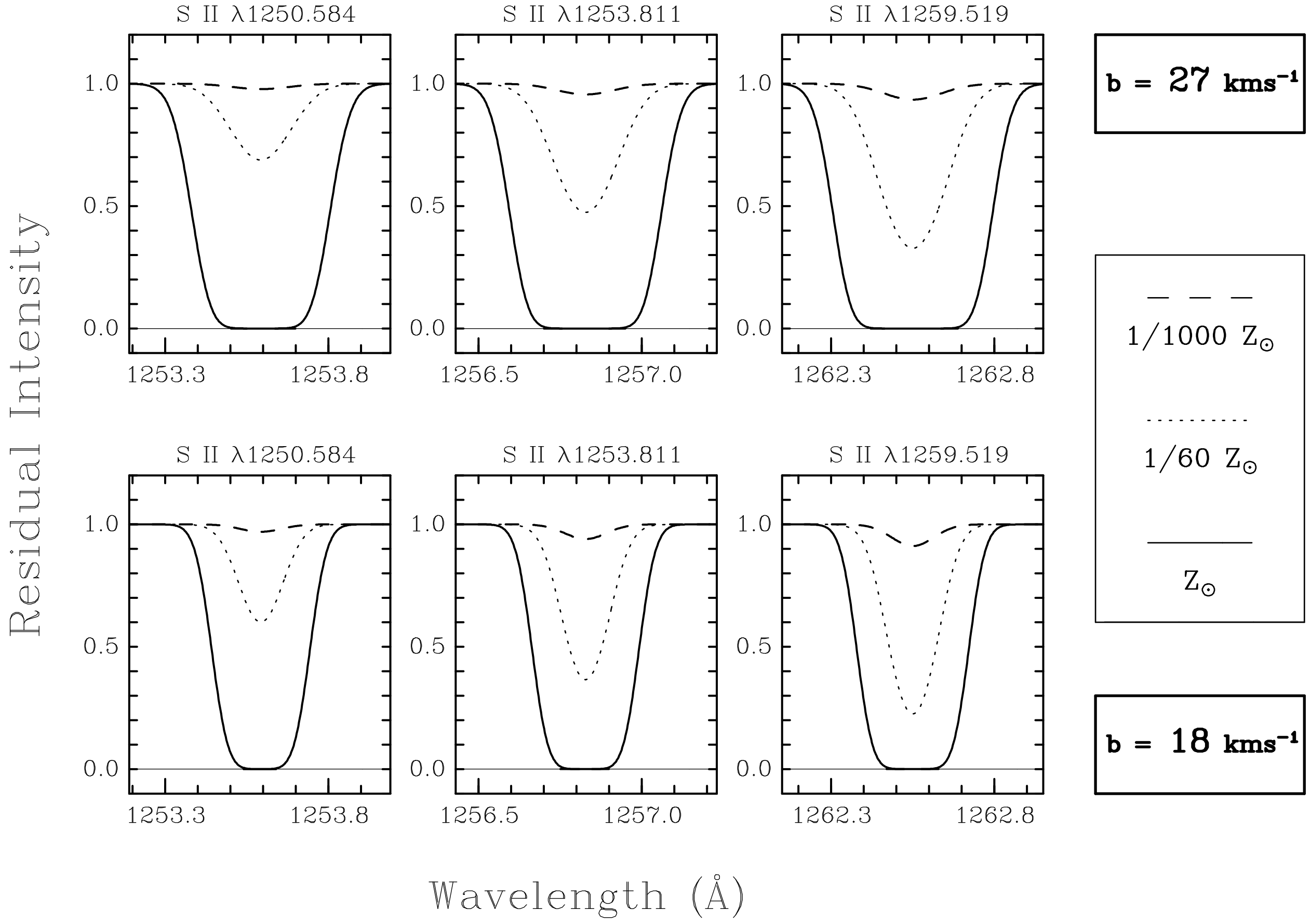

**Fig. 2.** Predicted profiles of the three lines of the S II λ1256.117 multiplet for three values of the abundance of S and two possible values of velocity dispersion $b$, as indicated. Future GHRS observations of these transitions with a moderate signal-to-noise ratio (S/N $\simeq$ 10) should be sufficient to measure the true metallicity of I Zw 18. The computed profiles have been convolved with the Line Spread Function of the repaired HST, corresponding to a resolving power $R = 16\,000$ (G160M grating).

## 3. Future Prospects

Is there no hope, therefore, of ever being able to measure the metallicity of the H I gas in I Zw 18? We briefly consider some possible strategies. For the reasons given above, it would not be profitable to increase the S/N of the existing observations. Even with the improved spectral resolution of the repaired HST, O I λ1302 is likely to be too saturated to be a useful diagnostic. On the other hand, λ1355 is too weak to be within reach. If [O/H] $=$ $-1.76$, as in the H II region, we expect $W_{1355} = 1$ mÅ; with a resolving power $R = 25\,000$, this would require S/N $\simeq 160$ to detect at the $3\sigma$ level. Higher O I multiplets occur at wavelengths too short to be accessed by HST.

In our view, the most profitable course of action is to target multiplet 1 of S II, λλ1250.584, 1253.811, 1259.519. S and O are both α-elements and are found in their solar relative proportions in H II regions of all metallicities, including I Zw 18 (Skillman & Kennicutt 1993; Garnett & Kennicutt 1994); it therefore seems reasonable to use S as a proxy for O as a measure of chemical enrichment. On the other hand, S is intrinsically $\sim 50$ times less abundant than O and the $f$-values of the lines in multiplet 1 are $3-9$ times smaller than that of OI $\lambda1302$ (Morton 1991). As a consequence, the strengths of the three S II lines depend much more sensitively on the S abundance, as can be seen from Figure 2. Evidently, S/N $\simeq 10$ should be adequate to assess if the metallicity of the H I gas is significantly below that of the H II region; furthermore, the fact that the $f$-values span a factor of 3 may well allow the degree of saturation to be ascertained and $N$(S II) to be deduced with some confidence.

While S II is the major ionization stage of S in H I clouds (only a negligible fraction of S is neutral), some of the S II could also occur in the H II region. However, an indication of the contribution by H II gas to the absorption is provided by the nearby fine structure line of Si II, $\lambda1264.7377$ which, being excited by electron collisions, is formed preferentially in ionised gas (Spitzer & Jenkins 1975). The GHRS spectrum obtained by KLSV and reproduced in Figure 1 required an integration time of 10 137 s to achieve S/N $\simeq 5$; thus a 40 000 s exposure centred near 1259 Å (the S II multiplet wavelength at the 740 km s$^{-1}$ redshift of I Zw 18) is probably sufficient to resolve the issue of the metal abundance of I Zw 18.

## 4. Conclusions

The hypothesis that the H II regions of dwarf galaxies are self-enriched in metals does encounter some difficulties. As Skillman & Kennicutt (1993) pointed out, it would then be coincidental that the two main emission regions in I Zw 18, separated by 5.6 arcseconds (or $\sim 270$ pc at a distance of 10 Mpc), exhibit very similar O abundances. Furthermore, in order to explain the relatively high N/O ratio in I Zw 18, only $\sim 5$ times below solar, it is necessary to invoke significant primary production of N from massive stars; current ideas, on the other hand, favour longer-lived intermediate-mass stars as the main producers of N at low metallicities (see the discussion by Kunth, Matteucci, & Marconi 1995). Yet, it will be necessary to confront these problems if the neutral interstellar medium is indeed found to be more metal-poor than the H II gas, as claimed by KLSV. We suggest that it may be prudent to defer such theoretical speculation until the basic observational result is placed on more secure ground.

We are very grateful to David Bowen for his kind help in retrieving data from the HST archive and for providing useful information on the analysis of GHRS spectra. MP acknowledges the generous hospitality of NORDITA, where this work was completed.